






\magnification1200
\baselineskip18pt plus 2pt minus 1pt

\newwrite\refout
\immediate\openout\refout=reference.txa
\newcount\countref  
\def\ref#1#2{\global\advance\countref by 1
\xdef#1{\number\countref}
\immediate\write\refout{\string\item{\number\countref.} #2}
}

\def\aref#1#2#3{
	\xdef#1{#3}
	\immediate\write\refout{\par\hangindent=.725
		truein \hangafter=1\ #2}
		}


\newcount\countsec

\def\newsec#1{\global\advance\countsec by 1
\goodbreak\medskip\goodbreak{\parindent0pt
{\bf \number\countsec. #1}}
\nobreak
\medskip
\nobreak\counteqn = 0
}
\def\Newsec#1{\global\advance\countsec by 1
\goodbreak\medskip\goodbreak{\parindent0pt
{\bf \number\countsec. #1}}
\nobreak
\medskip
\nobreak
}


\newcount\counteqn
\def\be#1{$$\global\advance\counteqn by 1
\xdef#1{(\number\counteqn)}}
\def\ee{\eqno(\number\counteqn)$$}
\def\ben{$$\global\advance\counteqn by 1}
\def\bea#1#2{\global\advance\counteqn by 1
\xdef#1{(\number\counteqn)}
             $$\eqalignno{#2 &(\number\counteqn)\cr}$$}
\def\bean#1{\global\advance\counteqn by 1
            $$\eqalignno{#1 &(\number\counteqn)\cr}$$}

\newwrite\figout
\immediate\openout\figout=figure.txa
\newcount\countfig
\def\putfig#1#2#3{
\global\advance\countfig by 1
\xdef#1{\number\countfig}
	\immediate\write\figout{\par\hangindent=.725 truein
\hangafter=1\ FIG.\ \number\countfig. #3}
}

\font\tenrm=cmr10 




\def\sts{\sigma_{\rm th}}
\def\sns{\sigma_{\rm n}}
\def\st{\sigma_{\rm th}^2}
\def\sn{\sigma_{\rm n}^2}

\def\done{d_1}
\def\dtwo{d_2}
\def\nc{N_c}
\def\np{N_p}
\def\lr{{\cal R}}

\def\ga{\mathrel{\mathpalette\fun >}}
\def\fun#1#2{\lower3.6pt\vbox{\baselineskip0pt\lineskip.9pt
  \ialign{$\mathsurround=0pt#1\hfil##\hfil$\crcr#2\crcr\sim\crcr}}}

\aref\BELM{Bond, J. R., Efstathiou, G., Lubin, P. M.,
$\&$ Meinhold, P. R. 1991,
{\it Phys. Rev. Lett.}, \ \ \ {\bf 66}, 2179.}{Bond et al. 1991}
\aref\BRANDT{Brandt, W.N., Lawrence, C.R., Readhead, A.C.S.,
Pakianathan, J.N.,
and Fiola, T.M. 1993, CalTech-OVRO preprint \# 12}
{Brandt et al. 1993}
\aref\DK{Dodelson, S. and Kosowsky, A., 1994, astro-ph/9410081.}
{Dodelson $\&$ Kosowsky 1994}
\aref\DST{Dodelson, S. $\&$ Stebbins, A., 1994, {\it
Ap. J.}, {\bf 433}, 440.}{Dodelson $\& Stebbins 1994}
\aref\Readhead{Readhead, A. C. S. {\it et al.}, 1989,
{\it Ap. J.}, {\bf 346},
566.}{Readhead et al. 1989}
\aref\saskatoon{Wollack, E.J., Jarosik, N.C., Netterfield, C.B.,
Page, L.A., $\&$ Wilkinson, D., 1993, Princeton preprint.}
{Wollack et al. 1993}

\font\titlefont=cmbx10 at 14.4truept

{
\nopagenumbers
\rightline{\hfill FERMILAB-Pub-94/039-A}
\rightline{\hfill astro-ph/9402053}
\rightline{\hfill February 1994}
\vskip 1.0truein
\titlefont
\centerline{Noise Correlations in}
\centerline{Cosmic Microwave Background Experiments}
\medskip\medskip\medskip
\noindent
\tenrm
\centerline{Scott Dodelson$^{1}$, Arthur Kosowsky$^{1,2}$, and Steven T.
Myers$^{3}$
}
\medskip
\centerline{$^1$NASA/Fermilab Astrophysics Center}
\centerline{Fermi National Accelerator Laboratory}
\centerline{P.O. Box 500, Batavia, IL 60510-0500}

\medskip
\centerline{$^2$Departments of Physics and of
Astronomy and Astrophysics}
\centerline{Enrico Fermi Institute}
\centerline{The University of Chicago}
\centerline{Chicago, IL 60637-1433}

\medskip
\centerline{$^3$Div. of Physics, Math, $\&$ Astronomy}
\centerline{California Institute of Technology}
\centerline{Pasadena, CA 91125}

\bigskip

\centerline{ABSTRACT}
Many analyses of microwave background experiments neglect the
correlation of noise in different frequency or polarization channels.
We show that these correlations, should they be present,
can lead to severe misinterpretation of an experiment.
In particular, correlated noise arising from either electronics or atmosphere
may mimic a cosmic signal.  We quantify how the likelihood function for a given
experiment varies with noise correlation, using both simple analytic models and
actual data. For a typical microwave background anisotropy experiment,
noise correlations at the level of 1\% of the overall noise can seriously {\it
reduce} the significance of a given detection.
\footnote{}{
Submitted for publication in {\it Astrophysical Journal Letters} }
\vfill\eject}

\pageno=1

\Newsec{Introduction}
The last few years have witnessed a surge of experiments measuring
anisotropies in the cosmic microwave background. The existence
of such
anisotropies is now firmly established;
measurements are becoming plentiful enough to compare
different cosmological theories quantitatively. A useful
interpretation of a given experiment relies on proper treatment
of atmospheric and instrumental noise. This {\it Letter}
\/focuses on one possible pitfall in analyzing experimental
results: noise in
different channels of an experiment may be
correlated. This correlated noise may mimic a cosmological signal on
the sky and significantly alter the interpretation of an
anisotropy measurement.

All present anisotropy experiments share several common features.
Typically, an experiment measures the deviation of the
microwave background temperature from its
mean value; this deviation is the temperature anisotropy.
Measurements are usually taken
at several different frequencies and sometimes at different
polarizations for a total of $\nc$ measurements at a given
point on the sky. This
set of measurements is then repeated in $\np$ patches on the sky.

The analysis of this type of experiment requires the correlation
function, which includes the expected contribution to the signal
from cosmological sources,
instrumental and atmospheric noise, and foreground sources. In
the present work, we ignore the last contribution; foreground
sources are considered in detail elsewhere (Brandt {\it et al.} 1993;
Dodelson and Stebbins, 1993; Dodelson and Kosowsky, 1994). The
correlation function and the data determine
the likelihood function (see, e.g. Readhead et al, 1989; Bond et al. 1991),
the probability of obtaining the data given a particular theory and the
noise parameters.
Explicitly, the likelihood function is given by
\ben
{\cal L} = {(2\pi)^{-N/2}\over\sqrt{ {\rm det} (C) } }
	\exp \left[  -{1\over2} D C^{-1} D \right],
\ee
where $N = \np\nc$ is the total number of data points,
$C$ is the $N\times N$ correlation matrix,
and $D$ is a $N$-component vector containing the data.
We are interested in small off-diagonal elements in the
noise contribution to the correlation matrix.

The actual value of the likelihood
function is not significant, but rather its relative value for
different potential theories. For the sake of simplicity,
we parametrize theories simply
by the variance they predict in a given experiment, $\sts$; then
${\cal L}$ is a function of $\sts$. A
maximum in the likelihood function at a non-zero value of
$\sts$ marks a detection; the
significance of a detection is reflected by the ratio of the
value of the likelihood function at its maximum to its value at
no theoretical signal. We therefore consider
the {\it likelihood ratio}:
\be\likerat
\lr(\sts)
\equiv { {\cal L}(\sts)
\over {\cal L}(\sts=0) }.
\ee

In this {\it Letter} we show analytically how the
likelihood ratio changes if small off-diagonal terms are
included in the correlation matrix. We conclude that
{\it even small off-diagonal correlations
can lead to huge changes in the likelihood ratio, and therefore
in the significance of a detection.} This analytical
work, in Sections II and III,
is useful but perhaps not completely convincing,
as it involves certain simplifying assumptions.
In Section IV we present
the likelihood ratio for the Saskatoon experiment (Wollack et al. 1993),
the only measurement of which we are currently aware
that reports off-diagonal correlations. The difference
between including such correlations in the analysis
(as the group properly did)
and neglecting them is shown to be dramatic. The Saskatoon
experiment, a ground-based apparatus which uses a single HEMT
amplifier for each three frequency channels,
has large noise correlations,
but even for experiments with substantially smaller
correlations the difference can still be very important.

\Newsec{Two Channel Experiment}

In this section we illustrate the importance of off-diagonal
correlations with a simple example. Consider an experiment
which measures the temperature anisotropies in two
frequency channels at one point on the sky. In the absence of
correlated noise, we need three pieces of information
to analyze
such an experiment: (i) the data, $D$,
which in this case consists of two numbers, the observed
temperature anisotropy in each channel; (ii) the theoretical
prediction for the expected rms anisotropy, $\sts$; and (iii)
the expected rms of the
noise, $\sns$. With two frequency channels, the latter
two quantities become $2\times2$ matrices.
The
correlation matrix
is the sum of these two matrices:
\ben
C_0 = \left\lbrack \matrix{
\st + \sn  & \st  \cr
\st  & \st + \sn  \cr}
\right\rbrack.
\ee
The theoretical rms $\st$ appears in every element, because
the expected rms due to the cosmic signal is the same
in every channel: if channel 1 measures a given signal $\done$,
channel 2 is predicted to measure the same value for $\dtwo$ in the
absence of noise. The theoretically expected signal in each channel
is therefore correlated. Any experiment will have diagonal
contributions to the noise; for simplicity we assume the same
noise rms in each frequency channel. Additional off-diagonal noise
components arise whenever the noise sources in different
frequency channels are correlated. We parametrize the
off-diagonal components by $\epsilon$ and
write the total correlation matrix as
\ben
C =\sn \left\lbrack \matrix{
1+x  & \epsilon+x  \cr
\epsilon +x  & 1+x  \cr}
\right\rbrack
\ee
where $x\equiv \st/\sn$.
Correlated noise will most likely arise from the atmosphere or from an
experiment's electronics.

We can evaluate directly the likelihood function in Eq.~(1) by noting
that
\ben
C^{-1} = {1\over \sn(1-\epsilon)(1+2x+\epsilon)}
\left\lbrack \matrix{
	1+x &  -\epsilon -x\cr
	-\epsilon -x & 1+x\cr} \right\rbrack.
\ee
Then writing the two measurements as $D\equiv\sigma_n
(\bar d +\done,\bar d+\dtwo)$, where $\bar d\sigma_n$ is the
mean measurement of the two channels, the
likelihood function is
\ben
{\cal L}={1\over 2\pi\sn}
\left[(1-\epsilon)(1+2x+\epsilon)\right]^{-1/2}
\exp\left[ -{\bar d^2\over 1+2x+\epsilon} -
{d_1^2 +d_2^2\over 2(1-\epsilon)}\right].
\ee
A straightforward calculation
shows that ${\cal L}(x)$ peaks at $x=X$ satisfying
\be\TMAX
2\bar d^2=1+2X+\epsilon
\ee
if $X$ is greater than zero.
This immediately gives the likelihood ratio, defined
in Eq. \likerat\ , as
\be\lrtwo
{\cal R} = \sqrt{ {1+\epsilon\over 2\bar d^2} }
		\exp\left\{ {\bar d^2\over 1+\epsilon}
				- {1\over 2} \right\}.
\ee
It is instructive to consider
a particular limit of Eq. \lrtwo. When the noise in each channel
is completely correlated ($\epsilon = 1$), we expect to get less
information from this experiment. Instead of two independent
channels, we really have only one independent channel. Thus the
likelihood ratio in this limit should be the same as for a
one channel experiment. A short calculation shows that for a one
channel experiment, the likelihood ratio is equal to
$(1/\bar d)\exp[ (2\bar d^2 -1)/2 ]$, which is indeed the value
of ${\cal R}$ in Eq. \lrtwo\ when $\epsilon=1$.

We now show that as long as the noise in the two channels
is positively correlated, the likelihood
ratio decreases. Let us define ${\cal R}_0$ to be
the likelihood ratio when noise is not correlated
($\epsilon=0$). Then, Eq. \lrtwo\ tells us that
\be\RMAX
{{\cal R}\over{\cal R}_0}=\sqrt{1+\epsilon}\exp\left[
-\bar d^2{\epsilon\over 1+\epsilon}\right].
\ee
This ratio is always less than one as long as
$0<\epsilon< 2\bar d^2 -1$. The first inequality
$(\epsilon >0)$ holds when the noise is positively
correlated; the second holds for any solution of Eq. \TMAX.
Thus {\it positively correlated noise reduces the likelihood ratio.}
This problem is particularly acute in cosmic microwave background
experiments, since the signal is also completely correlated in the
different channels; thus if the noise is correlated, we expect
the statistical significance of detections or upper limits to be
weakened.

Even for this simplistic one-patch, two-channel
model, the effect of noise correlations can be non-negligible
in estimating the significance of a detection. For example, if the
signal is twice the noise level ($\bar d^2 = 4$),
presently a representative signal-to-noise ratio,
and if $\epsilon \sim 0.5$ as it is in the Saskatoon experiment,
then including the correlation decreases the significance of a detection
by almost a factor of three. We will now show that the
situation gets
worse with multiple channels and patches.

\Newsec{Generalization}

It is straightforward to generalize the above discussion
to allow for many frequency and/or
polarization channels and many spatial patches.
For $N_c$ channels, $C$ becomes an $N_c\times N_c$ matrix with
components
\be\cbig
C_{ij}=\sn(1-\epsilon)\left[\delta_{ij} +{x+\epsilon\over 1-\epsilon}
\right].
\ee
Eq. \cbig\ idealizes an actual experiment in two ways:
(i) the noise in all the channels is assumed equal, so the
diagonal components of $C_0$ are equal; (ii)
each pair of channels has equal correlation, so the
off-diagonal components of $C_1$ are all equal. The inverse of $C$ is
easily obtained by assuming $C_{ij}^{-1}\propto \delta_{ij} + b$ and
imposing the condition $CC^{-1}=1$ to find the constant $b$ and the
normalization; the result is
\be\cinv
C^{-1}_{ij}={1\over\sn(1-\epsilon)}\left[
\delta_{ij}-{x+\epsilon\over 1+N_c x+(N_c-1)\epsilon}\right].
\ee
The determinant of $C$ is given by
\be\detc
{\rm det}\,C = \sigma_{\rm n}^{2\nc} (1-\epsilon)^{\nc-1}
\left[1+\nc x+(\nc-1)\epsilon\right],
\ee
which can be proven by induction. Combining these two expressions
gives the $\nc$-channel likelihood function:
\bea\nlike{
{\cal L} & ={1\over (2\pi)^{\nc/2}\sigma_n^{\nc}}
\left[(1-\epsilon)^{N_c-1}(1+N_cx+(N_c-1)\epsilon)\right]^{-1/2}
\cr
&\qquad\qquad\times
\exp\left[-{\nc\bar d^2/2\over 1+N_cx+(N_c-1)\epsilon} -
{1\over 2(1-\epsilon)}
\sum_i d_i^2\right],
}
which reduces to the previous result for $N_c=2$. The likelihood
function peaks at $x=X$ satisfying
\be\maxx
N_c\bar d^2 = 1+N_c X+(N_c-1)\epsilon,
\ee
and the likelihood ratio result generalizes to
\be\ratios
{{\cal R}\over{\cal R}_0}=\sqrt{1+(N_c-1)\epsilon}
\exp\left[-{\nc(\nc-1)\bar d^2\epsilon\over
2(1+(\nc-1)\epsilon)}\right].
\ee

A further generalization to $N_p$ different patches on the sky
can be approximated
by a block diagonal correlation matrix, where each block is an
identical $N_c$-channel correlation matrix. This approximation simply
raises the right side of Eq.~\ratios\ to the power $N_p$:
\be\ratiop
{{\cal R}\over{\cal R}_0}=\left[1+(N_c-1)\epsilon\right]^{N_p/2}
\exp\left[-{N_p \nc (\nc-1)\bar d^2\epsilon\over
2(1+(\nc-1)\epsilon)}\right].
\ee
where $\bar d^2$ now represents the mean squared signal-to-noise
ratio for all patches.
For the noise piece of the correlation matrix, this block diagonal
ansatz is normally a good
approximation, as noise in different patches is unlikely to be
correlated.
However taking the full correlation matrix
to be block diagonal is only an approximation, since
the signal is likely to be correlated from patch to patch unless
the patches are far removed from each other compared to the
patch size. The error in this
approximation does not qualitatively affect our arguments.

For a
typical medium angle experiment today, $\np \simeq
20$ and $\nc \ge 3$. With
a signal to noise ratio of order two, the argument of
the exponential
is typically greater than $200\epsilon/(1+2\epsilon)$.
If $\epsilon$ is of
order $0.5$, then
the significance of a detection decreases by a factor of
order $2^{10}e^{-50}
\simeq 10^{-19}$ when one accounts for the noise correlations.
For $\epsilon \simeq 0.1$, the significance decreases by $\simeq
10^{-7}$, and noise correlated even at the 1\% level may reduce the
likelihood ratio by nearly a factor of 10.
In general,
we expect correlations to be important roughly when
\ben
\epsilon \ga {1\over \np\nc(\nc-1)} .
\ee
As the signal-to-noise ratio increases, the effect of
correlations on the likelihood ratio becomes stronger,
although the likelihood itself becomes more sharply peaked.

\Newsec{Saskatoon Experiment}

In deriving the analytic result in Eq. \ratiop\ , we made three assumptions:
(1) The noise in each channel and patch was assumed to have the same
variance;
(2) The noise was assumed to be correlated in the same way
between any pair of channels;
(3) The signal was assumed to be uncorrelated from one spatial
patch to another.
In this section we analyze the Saskatoon
experiment without making any of these assumptions.
The reported error bars (the diagonal variance
of the noise) and the reported off-diagonal correlations
replace the first two assumptions. We perform the analysis
in the context of a standard cold dark matter (CDM) model
with a Harrison-Zel'dovich-Peebles initial spectrum, which
completely determines the correlations between patches.

The Saskatoon experiment takes measurements at six different
channels for each patch: three frequency channels and two polarization
channels. We consider the so-called ``East'' data set, which consists
of measurements in $21$ separate patches
(the ``West'' data set gives very similar answers). The theory gives the
predicted variance not only in a given patch, but also the correlations
between different patches. Specifically,
\ben
\langle d_a d_b \rangle  = \sum_{l=2}^{\infty}
	{2l+1\over 4\pi} C_l W_{l,ab}
\ee
where $a,b$ label different patches; $C_l$ is the prediction of
the theory for the $l^{\rm th}$ mulitple moment; and $W_{l,ab}$ is the
window function of the experiment which depends on the chopping strategy,
beam width, and spatial separation of patches $a$ and $b$.
For comparison, we previously assumed
$\langle d_a d_b \rangle = \delta_{ab} \st$.
CDM has only one free
parameter, so $C_l/C_2$ is fixed for all $l>2$.
Figure 1 shows the likelihood ratio $\lr$ for the Saskatoon East
data as a function of
$Q_{\rm rms}=\sqrt{5C_2/4\pi}$.

The two curves in Figure 1 correspond to the likelihood ratio
with and without off-diagonal noise correlations. The difference
is stunning. A detection which would have been extremely clean
$[\lr(\bar C_2) \sim 10^{11}]$ becomes much less certain
$[\lr(\bar C_2) \sim 30]$ once correlations are accounted for.
(Note that Eq. \ratiop\ actually
underestimates the effects of noise correlations
in this case: $\st/\sn = .68$ and $\epsilon \sim 1/3$,
so that we expect ${\cal R}/{\cal R}_0 \sim 5\times 10^{-8}$.
One reason for this is that
 the off-diagonal noise elements
were chosen to be equal in the simple model leading
up to Eq. \ratiop\ , whereas in any real experiment, certain
channels will be more strongly correlated than others.)
We emphasize that the Saskatoon experiment
{\it did}\/ include this effect in
their analysis; we use this experiment as an illustration
because it is the {\it only}\/ one
of which we are aware that has reported the presence (or lack of)
noise correlations.

In summary, even relatively small amounts of correlated noise between
frequency channels can greatly affect the interpretation of
microwave background experiments. The effect is likely to be of
particular importance for those experiments which utilize a common
part of the electronic signal path for multiple frequency channels
(as in the Saskatoon experiment considered here), or for experiments
with substantial atmospheric or environmental sensitivity.
We hope
the arguments presented here will prompt other groups to examine and
report noise correlations in their experiments.

We are especially grateful to Lyman Page and Stephan Meyer for
their patient explanation of the Saskatoon experiment, and
to them and Michael Turner for other useful discussions.
This work was supported in part by the DOE (at Chicago and Fermilab)
and by NASA through
grant No.
NAGW-2381 (at Fermilab). AK is supported in part by the NASA Graduate
Student
Researchers Program.

\immediate\closeout\refout
	\medskip
	\centerline{\bf REFERENCES}
	\medskip
	\input reference.txa
\bigskip
\centerline{\bf Figure Caption}
	\medskip
\noindent
Figure 1. The Likelihood Ratio vs. $Q_{\rm rms}$ for the East
data of the Saskatoon experiment with and without noise correlations.
$Q_{\rm rms}$ is related to $C_2$ via $Q_{\rm rms}^2 = 5C_2/4\pi$.

\bye